# Automating Truth: The Case for Crowd-Powered Scientific Investigation in Economics

Jorge M. Faleiro Jr [#1]

*Centre of Computational Finance and Economic Agents, University of Essex*
*Wivenhoe Park, Colchester, CO4 3SQ, UK*
[1] jfalei@essex.ac.uk   j@falei.ro

*Abstract*[1] — **Scientific investigation procedures have been evolving to follow an ever-changing cultural landscape, the sophistication of the technology available and an ever-growing knowledge base. This continuous evolution brought investigation practices through distinct historical phases, mostly marked by different types of participants and organization, from individual natural philosophers to science driven by large institutions.**

**There is clear evidence that we are now getting to an age of drastic disruptive change. Increased complexity and mandatory multidisciplinary thinking have moved research from an initial phase of disjoint polymaths into a current phase of widespread uncontrolled use of computational tools and data generation, the "informatics crisis". The use of advanced computational technology for communication and generation of data in large scale without proper controls is compromising our ability to conduct an adequate reproducible investigation, causing a dangerous drift from the scientific method. The same technology that could potentially control and automate the production and analysis of results is undermining the principles of the scientific method.**

**To counteract this deviation, we advocate the use of a next-generation investigative approach leveraging forces of human diversity, micro-specialized crowds and proper computer-assisted control methods associated with a "pipeline of proof".**

**This paper outlines the impact of advanced computational technology, not only as an accelerator of the rate in which humanity acquires objective knowledge but also as a dangerous side effect as a generator of massive amounts of uncontrolled, unverified and untraceable data and results that cannot be reproduced.**

**We propose an alternative for methods of investigation based on collaboration in large-scale through standard procedures of proof and crowds in building a "collective brain in which neurons are human collaborators".**

## I. Scientific Learning and Economics

Humans learn new things through investigation. Careful investigation is what establishes if an observed phenomenon is real, or it should be deemed just a result of random forces of nature at play. The primary target of any investigation is to establish facts, as accurately as possible, by proving observations to be either true or false. That is how humankind has been accumulating objective knowledge for as long as we walk this earth, and this is why defining precise methods of proof is crucial.

However, the mental process we follow as individuals to investigate and learn about things is not straightforward. Even at this present date, science is still not able to unequivocally explain the process by which we learn and assess things. If this is true when we produce our thoughts on our own, we should expect an even more elaborate process to be at play when we introduce procedures of investigation that are performed by multiple individuals, organized in seemingly chaotic crowds.

The process of learning and understanding followed by humans is abstract, fluid and subject to multiple definitions of what one might consider knowledge, assumptions, and beliefs. Given this abstract nature, it is essential to identify what is objective knowledge, or what is known, from what is not. Alternatively, in other words, demarcate the difference between what is considered science from what is not.

The clear demarcation of what is considered science, and what is not, is part of a controversial issue usually referred to as a *demarcation problem* [1]. In a scenario where the intent is to produce new knowledge from a pre-existent foundation of knowledge by using large crowds of participants in scientific investigation, it becomes even more critical establish clear criteria for demarcation and for separation of what is known from everything else. This process of building new knowledge from a pre-existing foundation of what is considered to be true is called *scientific learning*. Scientific learning occurs as a result of two specific requirements.

The first requirement is that by definition scientific learning occurs by the application of the principles of the scientific method. These principles show that, despite its power, science is indeed a simple tool. In science, we rule out things considered false based on hard evidence. What is true (truthfulness) is then inferred by exclusion[2] [2] [3] [4]. We refer to this process of inference as the *modern scientific method*, outlined by a set of six principles [5]:

- The goal of scientific investigation should be to gain objective knowledge [6].
- Scientific knowledge is obtained through tests, experiments and observations. Tentative assumptions

---
[1] Large portions of this paper are reproduced as part of [21]

[2] Discounting the recent resurgence of the truth-conducive controversy, in which "it is fashionable among (…) some philosophers to say there are no principles of rationality that are truth-conducive (…) since there are no standards of rationality, there is no logic or method to science" [111].

about a particular phenomenon may, however, be deduced from pre-existing knowledge.

- A hypothesis must be verifiable by some experimental or observational method.

- Experiments must be reproducible and must have controls

- The integrity of the data must be appropriately safeguarded.

In the modern scientific method, "each principle helps to increase the reliability and accuracy of knowledge resulting from scientific research" [5]. By that definition, these principles naturally address the requirements for objective scientific learning in economics described previously.

The second requirement is that scientific learning is dependent on peculiar characteristics in a field of study. Economics and financial sciences[3] are particular domains of knowledge in which related systems and agents – markets, humans and their relationships – are hard, if not impossible, to model. For appropriate investigative procedures, adequate financial models must be able to deal with this intrinsic complexity of economic systems and agents [7] [8] [9]. As such, economics should not be treated differently from other disciplines considered hard sciences [4] [5]. In the field of economics the process by which knowledge is acquired is defined, and dependent, on three specific peculiarities of the subject of study:

- **Complexity**: modern economics deals with a unique subject of study - a shared, intertwined, complex market – that cannot be rewound. Time like life moves towards one direction [10]. Given the usually large number of inputs to such a complex system and the apparent independence between these input variables, once an event occurs, we cannot derive different futures from what the present currently describes [6].

- **Lack of proper theoretical models**: when taken from a recent historical perspective modern economics has been associated with compartmented classical fields like psychology, statistics, sociology, and computer sciences. Most of the assumptions in classical and theoretical sciences are inherently oversimplified and flawed when trying to predict or understand the behavior of a systemic market[6] [10] [6].

- **Multidisciplinary fields of study**: modern economics is, in essence, a multidisciplinary subject. Efforts to understand the market considering its most fundamental structures tend to rely on somewhat orthogonal fields of study like neuroeconomics [11], behavioral sciences [12], and analysis of market micro-events [13], among others. The interdependency of subjects in economics to bioengineering, neurosciences, social sciences, psychology, data and computer sciences, and related fields is diffuse and difficult to correlate and at the same time, critical for scientific learning [6].

As a consequence of these peculiarities of our field of study (i.e., systemic complexity, lack of proper theoretical models, and novelty of correlated fields of study) modern research in economics becomes strictly dependent on high-performance computers [7], requiring the implementation of elaborate simulation-based techniques. Similar to what is used in other hard-sciences, such as physics, engineering, and biophysics [14] [15] [6]. This dependency on high-performance computing has driven research in economics to favor specialized techniques for storage and processing speed. The field has been shaped so that the sheer generation of data and obscure ways to represent computational procedures is prioritized over proper control.

To offset these limitations, this paper advocates the use of crowds for the investigation and resolution of complex problems in general and in economics in particular, an approach we are calling *crowd-based investigation*.

## II. CROWD-BASED SCIENTIFIC INVESTIGATION AND A PIPELINE OF PROOF

Given the number of participants and the nature of the interaction – formal scientific investigation - we can safely expect as a consequence a large number of hypotheses being generated and tested. On this scenario, ideas must be defined, exchanged, discussed, and tested in a sequence of steps, arranged like a pipeline. Procedures in each step of the pipeline should potentially generate massive amounts of data, and each piece of data should be unquestionably tested as true or false. As a consequence, each of the steps must abide by transparency standards and validation metrics that must be well understood and accepted by all participants.

In this section, we define requisites and the composition of this process and steps involved, what we call a *proof pipeline*.

A proof pipeline for scientific investigation is proposed as a process, composed of individual tasks, each task producing standard outputs that can be used to prove or reject an observation.

A simplistic description of such a pipeline would be a tube, where its input, taken on the head of the pipe, is a problem, or a set of ideas under investigation, and other intangible aspects such as the experience of the individual performing the inquiry or the investigation. On the tail of the tube, the result of the investigation, as either true or false. Over the extension of the tube, there are small holes, from where the process produces pre-defined, controlled data as evidence. A diagram of a proof pipeline is shown later in this thesis, in Figure 1, on page 4.

---

[3] In the context of this document the terms "financial sciences" and "economics" have interchangeable connotations.

[4] We use the term "hard sciences" as it was coined by Nobel Prize winner in Economics in 1978 Herbert Simon, "for his pioneering research into the decision-making process within economic organizations", on his words: "The social sciences, I thought, needed the same kind of rigor and the same mathematical underpinnings that had made the *'hard' sciences* so brilliantly successful" (Simon 1978)

[5] We incorporate the colloquial definition of "hard" and "soft" sciences to respectively discern between natural sciences (e.g., biology, chemistry, and physics) and social sciences (e.g., economics, psychology, sociology) based on "evidence of a hierarchy of sciences" (Fanelli and Glanzel 2013).

[6] While we consider important to highlight this peculiarity, evaluating reasons for such limitations, or trying to entirely refute or confirm them is beyond the scope of this research

[7] A correlated consequence is that an ever-increasing dependency on high performance computers for scientific investigation makes it difficult to differentiate between subjects that are specific to economics, financial sciences, or computational finance. In other words, there is an incentive and a justification for economics, financial sciences, and computational finance to have a significant overlap.

The idea of arranging a sequence of pre-defined steps to assert a result of an investigation as true or false is not new. There are references in the literature to a step-by-step process in biomedical research, specifically for statistical measurements, referred to as a "statistical pipeline" [16] [17]. Although similar in its overreaching purpose and the intended standardization of the understanding of what is true or false, the scope of what that pipeline would entail is different than what this paper proposes. Their scope is also limited specifically to software patterns and a computational platform. Specifically, in the field of economics, there are proposals in the literature with minor overlapping with the idea of proof pipelines, arranging economic models as testable pieces of engineering, not necessarily as pipelines, referred to as "economic wind tunnels" [14].

As described earlier, a proof pipeline is a process, and as it is usually the case with processes, each step or part is composed of smaller mechanisms, smaller gears. Some gears are familiar and well understood, others not so much. One of those gears, required to establish a proof pipeline, is the underlying mechanism by which we get to conclusions based on premises taken from specific outcomes of an investigation. This process of getting to conclusions based on premises is called *inference*[8]. An inference is a mechanism we use to evaluate, learn, and create. This intricate mechanism is responsible for some of the most fundamental structures of the scientific thought.

The mechanism of inference is a complex and abstract subject, and for its very nature, it is difficult to explain. Human ingenuity is attracted to things that can't be easily explained, so scientists have been looking at the general subject of inference for a long time, trying to understand and explain the specifics through studies in philosophy, biomedicine, and even religion. This lengthy inquiry is far from over. Formalizations of the exact mechanisms at play are mostly abstract, and as it is usually the case with philosophical subjects, surrounded by controversy [18] [19].

For this reason, in this paper, we want to carefully, and intentionally, stay away from the argument. While we understand the importance of the debate and study of general concepts around the topic called "philosophy of science" [20], each of the small topics under the subject could lend a lengthy separate study in itself. Would be impractical and redundant to explore in this paper all the open controversies, different viewpoints, intricate details, and differences between methods [21].

Hence, it is essential at this point to carefully define our scope of interest when it comes to the general topic of inference and a proof pipeline, namely four specific topics:

- **Support for falsifiable and testable inquiry**: the demarcation of what should be considered scientific is given by investigation propositions formalized by statements that can be tested and falsified. A scientific statement should be capable of conflicting with possible or conceivable observations, in line with the principle of falsifiability, in which "statements or systems of statements, in order to be ranked as scientific, must be capable of conflicting with possible, or conceivable observations" [22]
- **Step-wise, algorithmic nature**: methods of inference should fit a general algorithmic structure and a step-by-step, procedural description, mimicking the sequential arrangement of a pipeline.
- **Participation and collaboration in large-scale**: investigation should incentivize collaboration and interaction of a large number of participants. Features or metrics of inference should be well understood and serve as a quantifiable standard for what to be considered true or false.
- **Computer augmented**: computers should serve as control points for collaboration and interaction of human participants, and not as agents of scientific inquiry themselves.

Given the scope of inference and these topics, the most commonly accepted model of inference describing the scientific inquiry based on testing and falsifiability is the *Hypothetico-Deductive model*, or *H-D model* [23]. The H-D model is a composition of all known modes of reasoning [24] [25] [26]. In this sense, reasoning is defined as the act of associating premises to conclusions and is described through three distinct modes of reasoning: deductive, inductive, and abductive. These modes of reasoning are considered the core of Karl Popper's falsifiability and testability argument of any scientific hypothesis [27].

Popper's surprisingly simple theory proposes discovery to occur in two steps. On the first step – *conjecture*[9] – a scientist offers a hypothesis that might explain some natural phenomena. The second step – *refutation* – the hypothesis is tested in order to show that the hypothesis is false [22]. If we succeed to show that the original conjecture is false, we go back to the first step, build a new conjecture, and follow the two-step process again. If in the second step we fail to test the hypothesis as false we should assume that the original conjecture is – for the moment, and as far as we could not prove otherwise – correct [20].

Popper's theory is fundamental to the definition of the proof pipeline proposed in this section through a variation of the H-D model, and application of all modes of reasoning. The exact formulation of the H-D model vary, but in most cases, it is a combination of Karl Popper's view of falsifiability and testing, and a less skeptical view about confirmation[10] [20]. A less skeptical view, in this case, means that our reliance on the notion that evidence can affect the credibility of a hypothesis is necessarily fallible[11] [28].

---

[8] In this paper the term *inference*, used without qualifications, refers specifically to *human inference* and is defined as "the act of passing from one's proposition, statement, or judgment, considered as true, to another whose truth is believed to follow from that of the former" (Merriam-Webster 2018). Inference performed by humans and machines are related to different mechanisms and should not be used interchangeably (Gellatly 1989). It also differs for the term *statistical inference*, or *quantitative inference*, also used in this paper, defined as "the act of passing from statistical sample data to generalizations usually with calculated degrees of certainty" (Merriam-Webster 2018).

[9] A conjecture is not materialized as a specific contribution. As a consequence, multiple experts can express the same semantic, and therefore one conjecture can possibly be reflected in different models.

The essence of the idea behind the hypothetic-deductivism in science is old, with its origins in Plato's dialogues, referred to in that work as "the method of hypotheses" [23]. In a broader sense, the H-D model relies on a proposition of a hypothesis in a way that it can be falsified by a test of this proposition against observations, or evidence. The H-D model represents a formalization of the scientific method through a set of a simple sequence of four steps: observe; form a conjecture; deduce predictions from a conjecture; and test the predictions [20].

Additionally, the H-D model formalizes a process of investigation through individual, sequential steps. The formalization of a process of investigation through a pre-defined sequence of steps defines the process of discovery as inherently algorithmic [29]. The idea of algorithmic procedures of investigation is not new. A precursor of the modern scientific method, Francis Bacon, arguably a predecessor of Karl Popper in respect of the method of falsification [30] had foreseen two critical interconnected insights that are relevant to this research:

- The step-by-step, methodical approach to investigation, where Bacon used the word "machine" to describe his method in *Novum Organum* in 1620 [29] [30] [31].
- Bacon's method intended to leverage a "community of observers to collect vast amounts of information" and tabulate it into a central repository accessible to all [29], what would be equivalent to the notion of a what today we call a crowd in XVII century parlance.

Following through on Bacon's hint, if discovery is algorithmic, then we can safely assume that machines could perform it. Alternatively, better yet, as this paper advocates, discovery can be performed in large scale, having machines orchestrate the steps and rules of the collaboration of human crowds.

The application of this H-D model as an algorithm to a framework supporting crowd-based investigation can be described through a set of specific steps [20]:

- **Observe**. The observer should use personal experience to understand and appreciate the problem under study. Gather previous contributions[12] relevant to the case of use at hand.
- **Form a conjecture or hypothesis** [13]. Form a supposition, or a proposed explanation for the phenomena under observation, based on whatever limited evidence has been currently gathered as a starting point for further investigation. State an explanation of the hypothesis. Materialize that conjecture as a model. Share that model.
- **Deduce predictions from the conjecture**. Formalize predictions, stating what should be expected if the conjecture is true. Incorporate those predictions as part of the model.
- **Test**. Experiment with the model, looking for evidence (observations) that might disprove the predictions. Record all evidence as contributions and share those contributions. If predictions are disproved, so is the hypothesis: go back to step 2 and repeat.

This sequence of steps in a pipeline of proof is depicted in Figure 1.

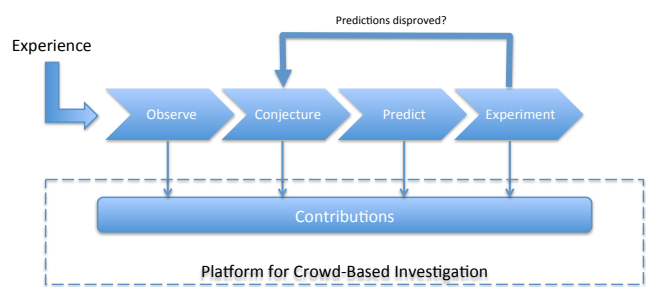

**Figure 1. Method of Proof in Crowd-Based Investigation**

A method of proof for collaboration in large-scale applying variations of the Hypothetico-Deductive Model to handle shared procedures of investigation in each of the phases: observe, conjecture, predict, and experiment.

This paper advocates the use of this process, created from features of the H-D model, as a baseline for a proof pipeline for a crowd-based investigation and validation through the exchange of shared evidence.

*A  Observation*

The first step of a proof pipeline deals with human *observations*. Observations are organoleptic and by definition are subject to abstract human interpretation. Given its subjective nature, it would be hard, if not impossible, to use machines to automate the process by which we generate high quality, reliable observation records. On the other hand, machines should be ideal to establish a platform for collaboration in large-scale where observations can be recorded and shared.

---

[10] Confirmation refers to "the problem of understanding when observations can confirm a scientific theory", and what is required in order to have an "observation evidence for the theory". This is a complex philosophical problem, often referred to as "the mother of all problems" [20].

[11] Observations cannot confirm theories or conclusions, i.e., "even with extensive and truthful evidence available, drawing a mistaken conclusion is more than a mere possibility", and as a consequence "under usual circumstances, reasoning from evidence is necessarily fallible" (Crupi 2016)

[12] For now, the term "contribution" here is used in the same sense as when authors "contribute" to Wikipedia. A more detailed formalization of contributions is done in separate publication [21] [114].

[13] Literature refers to the specific step in the H-D model where a supposition or specific explanation is made as a *conjecture*, which is equivalent to the most commonly known term *hypothesis*. For fairness, and accuracy, this paper refers to both terms interchangeably.

The idea of collecting observations in large-scale is not new. In *Novum Organum,* in 1620, Francis Bacon proposed a method intended to leverage a "community of observers to collect vast amounts of information" and tabulate them into a central repository accessible to all [29]. Currently, modern technology allows that vision of a central registry of contributions that can be shared and evaluated by a community of observers.

### B  Conjecture

The second step of the proof pipeline is the generation of *conjectures*, or *hypotheses*, to attempt to explain the cause of phenomena under observation. Hypotheses must be falsifiable, and at the same time, by definition cannot be entirely and irrefutably confirmed. It should always be assumed that improved research methods should disprove a hypothesis at a later date.

Assuming an algorithmic nature of the discovery process, in what is commonly called *automated hypothesis generation*, hypotheses have a potential to be entirely generated by computers[14] [32]. Automated hypothesis generation is still in initial stages, but research has produced a significant number of exceptional use cases.

Starting in the 1980's some experiments were able to hypothesize links between cause and effect, in two initially unrelated fields of study, without specialized knowledge in any of the subjects of study, and without conducting any experiments. The links were established by merely following algorithmic steps while connecting scientific papers with no citation overlaps [33]. More recent research allows for limited automated hypothesis generation based on large-scale text mining of academic publications, natural language processing, mathematical modeling, and graph theory. Some equally relevant and related features, taken out of the techniques in use in hypothesis generation, include the prediction of a successful academic career based on the writing style of scientists on entry-level positions and quantifiable metrics of efficiency in scientific discovery [34] [35] [36] [37] [32].

However, despite the evidence of potential progress and the slow advancement, science still lacks a complete theory for fully automated hypothesis generation. Additionally, these techniques currently rely on volumes of quality data associated with scientific publications, a scarce resource now that major scientific journals have placed severe restrictions on text mining of their content [38].

Instead of a fully automated hypothesis generation, this paper advocates the use of highly interconnected crowds orchestrated by computers. In such an environment, individual participants in a crowd would rely on computers to perform specialized discovery tasks, communication, and to produce metrics of quality on shareable contributions. As a consequence, hypotheses are generated by participants in a crowd, in an environment enhanced by computers, and not solely performed by machines.

### C  Prediction

The third step of the proof pipeline relates to *prediction*, where a researcher generates anticipations of probable outcomes of experimentations assuming that initial conjectures produced in the previous step, described in Section B, are true.

The mechanisms used to generate valuable, and high-quality predictions are similar to mechanisms we use to anticipate and track patterns in experience [20]. These mechanisms are subject to the complex rules that govern the connection of experiences, or the rules of science itself[15] [39] [40]. These complex rules are bound to human traits of creativity and experience and, as of the time of this writing, there are no instances of efficient implementation in machines.

In the same manner, probable outcomes can be defined as different shocks of executions [21]. Shocks are by definition an iteration of a simulation. The results of the execution of individual shocks are recorded in datasets as shareable evidence, allowing other participants to understand the expectations of a model better and assess predictions against actual outcomes.

### D  Test

The last step of the proof pipeline is *testing*, where experiments are designed based on predictions produced during the Prediction step, described in Section C. Those experiments are performed in order to support or refute predictions, and the outcome of a test would either validate or falsify the original conjecture, or hypothesis.

In some fields of study reliant on intensive and controlled testing, procedures related to experimentation are widely automated. Scientists can submit a description of their experiments online and have that description subsequently converted to specialized instructions and fed into robotic platforms to execute a battery of repeatable experiments [41] [29].

If we are to consider the assumption of standardized, quantifiable, and normalized results, it is important to introduce at this point the notion that experimentation on a complete method should also incorporate probabilities. In this case, a prediction should be expected to hold true *N*% of the time, in which case experimentation should be repeated to substantiate the probability *N* [42].

On this sense, achieving unambiguous conclusions about a problem then becomes a numerical exercise, in which statistical inference[16] is the process of getting to conclusions about a specific problem by looking at statistical characteristics of data, and by using probability alone [18] [43].

---

[14] There are notable exceptions to the belief that discovery can be algorithmic. "Karl Popper insists there is no recipe for coming up with interesting conjectures" (Godfrey-Smith 2003)

[15] The core objective of science is to understand how experience shapes discovery, on the words of Moritz Schlick "what every scientist seeks (…) are the rules which govern the connection of experiences, and by which alone they can be predicted" (Mulder and van de Velde-Schlick 1979b)

[16] The term *statistical inference*, or alternatively *quantitative inference*, is defined in this paper as "the act of passing from statistical sample data to generalizations usually with calculated degrees of certainty" (Merriam-Webster 2018).

There is a widespread agreement that statistics depend on probability, but concomitantly there are disagreements as to what exactly is probability, and how probability is connected to statistics[17] [44]. Over the last several decades Ronald Fisher [45], Harold Jeffreys [46], Jerzy Neyman [47], Leonard Savage [44], and many of their followers have defined several paradigms and have engaged in a number of debates that gave birth to controversies that were key to its formative properties [48]. A positive and possibly unintended consequence of the debate is the multitude of statistical tools and the rich set of options available to the scientific community to conduct quantitative inference [49] [50].

On this research, we acknowledge that these differences are essential, but we assume that even more important is to leverage this toolset to concentrate on relationships between data and model, or how representations mapping measurements in the real to the theoretical world are made. This shift in paradigm, in which statistical models take a back seat to the understanding of the relationships between data and methods to infer conclusions, is called *statistical pragmatism* [49]. In statistical pragmatism numerical methods are seen as an eclectic practice, emphasizing mechanisms by which observed data is connected to statistical procedures, as described in Figure 2 [49].

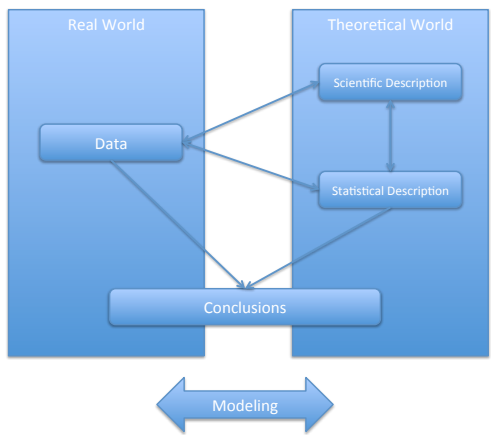

**Figure 2. Pragmatic Statistics and the Mapping Between Data and Methods**

Pragmatic statistics are defined through abstract mathematics constructs used to quantify and explain observable phenomena [49].

In essence, pragmatic statistics is a vehicle to aggregate real data and theoretical descriptions into quantifiable results, and as such can be seen as a model to reach a set of conclusions based on real and theoretical constraints.

---

[17] The definition of probability is at the root of the division on the understanding of what is physical and evidential probability, as "it is unanimously agreed that statistics depends somehow on probability. But, as to what probability is and how it is connected with statistics, there has seldom been such complete disagreement and breakdown of communication since the Tower of Babel. Doubtless, much of the disagreement is merely terminological and would disappear under sufficiently sharp analysis." [44]

## III. REQUIREMENTS FOR CROWD-BASED INVESTIGATION

The use of crowds for resolution of problems follows one of two distinct approaches.

The first approach, named "wise crowds" [51] relies on empirical observations [52] [53] and assumes the existence of some invisible, unquantifiable mechanism, somehow providing a certain level of knowledge to crowds, and therefore allowing them to make wise decisions. The "wise crowd" approach relies on the assumption of complete independence and isolation between participants of a crowd.

The second approach, named *collaborative crowds*, assumes that knowledge is produced as a result of structured collaboration between participants of a crowd.

This research subscribes to the second approach, collaborative crowds, where large-scale collaboration occurs by the existence of particular requirements of collaboration, as a natural evolutionary response to the environment in which investigation takes place.

The use of crowds as agents of investigation requires an organization of considerable number of individuals, in different roles and at different levels of technical understanding, to continually collaborate for the resolution of complex problems. However, as we can readily ascertain by observation, collaboration does not come out of thin air. We need something to drive effective collaboration, and in this section, we concentrate on explaining what "that something" is: the requirements for effective collaboration to take place.

Collaboration is what builds "some sort of a collective brain with the people in the group playing the role of neurons" [54] [51] and ultimately amplifies the intelligence of a group of people. Collaboration is facilitated as a result of four *requirements*: expert attention, proper cultural and intellectual development, manufactured serendipity, and human diversity.

- **Expert Attention:** Maximizing collaboration is primarily a problem of restructuring expert attention by designing the correct incentives that would encourage any single participant in a crowd to play the role of an expert at times, whenever it is required. Given the over-specialized nature knowledge and the narrow window of expertize, these experts in crowds are called micro-experts. Being able to call "the attention of the right expert at the right time" is critical to problem resolution by crowds of individuals. "Expert attention is to creative problem solving what water is to life: it's the fundamental scarce resource" [54].

- **Proper Cultural and Intellectual Development:** Collaboration must rely on participants in a proper stage of cultural and intellectual development. In the upcoming Section V, we describe the historical evolution of the scientific process: from individual macro-experts to institutionalized science to the anticipated, next phase of a crowd-based investigation. As part of this evolution, we start to notice evidence of disruption in the current discovery process based on hierarchies and institutions, and the transition to a new phase in which discovery is driven by crowds and micro-experts. This disruption and transition are discussed in details in the upcoming Section VI.

- **Manufactured Serendipity:** Collaboration requires the right participant with the ideal amount of micro expertise to help in the resolution of a problem. This phenomenon called manufactured serendipity, allow for fortunate discoveries of possible opportunities for collaboration by quasi-accident. Serendipitous connections between individuals are known to be essential in any creative or investigative work. The thinking behind "manufactured serendipitous connections" [55] assumes connections between individuals cannot be fabricated, but the conditions by which they occur can be stimulated on purpose. In other words, "you can't automate accidental discoveries, but you can manufacture the conditions in which such events are more likely to occur" [55].
- **Cognitive Diversity:** The last requirement calls for cognitive diversity. Cognitive diversity is the "extent to which a group of people reflects differences in knowledge, including abstract constructs like beliefs, preferences, and perspectives" [56]. Collaboration groups must be cognitively diverse, so to maximize instances of micro-expertise amongst its participants [57]. Putting it differently, to maximize collaboration, participants need a wide range of non-overlapping expertise. The minimum amount of shared knowledge must be the level that would allow participants to communicate effectively [51] [54].

These requirements reflect the need for collaboration in investigative procedures that often are cross-disciplinary. Existing literature has identified by empirical methods a different set of requirements, but in a perspective that seems influenced by thinking on that specific field of study (e.g., in social sciences these same requirements for collaboration are outlined as process, understanding, utility, and knowledge integration [58]). As a common limitation, other instances in the literature lack quantitative metrics to show evidence of the importance of each prospective feature in collaboration. We call *collaboration metrics* the quantification of requirements for large-scale collaboration [21].

IV. DEALING WITH COMPLEXITY: COLLABORATIVE RESOLUTION OF COMPLEX PROBLEMS

As described in Section I, this paper assumes that organized human collaboration is well suited for the investigation and resolution of complex problems. In reality, it would be impossible to infer absolute suitability of large-scale collaboration for the resolution of complex problems. Alternatively, we can enumerate results from empirical exercises, and their specific details, as evidence of resolution of complex problems by crowds.

The first example, the Polymath Project [59] is a brainchild of Fields Medal winner Timothy Gowers, a mathematician at Cambridge University. The Polymath project started with a pair of simple posts on his blog. The first post inquiring on the possibility of the use of crowds in the resolution of complex mathematics problems [60], and then shortly after that a second post where Gowers proposed a particular problem to be resolved using massively collaborative investigation [61].

Over the next 37 days, 27 people from around the globe – from mathematics enthusiasts to high school math teachers, and other Fields Medal winner Terence Tao – wrote 800 comments and more than 170,000 words on erratic movements of discovery [54] following an open path of investigative try-and-error. After those 37 days, Gowers announced that the crowd had solved not only the original problem but also a harder, more generic problem that had the initially proposed problem as a special case [62][18].

Considering the requirements for large-scale collaboration introduced in Section II, the original problem was not proposed on the most appropriate platform for collaboration – basically, a sequence of textual comments on Gower's online blog – and the specifics of the methods of incentive for micro-specialists was not clear. Despite that, over the following months around a dozen of unresolved problems were proposed and resolved by a crowd of mathematics investigators, and the platform was moved from an online blog to a wiki [63]. Despite lacking an adequate computational representation for investigation in mathematics, the Polymath Project is a successful example of investigation and resolution of specialized, very complex problems by crowds.

The second example is the control of predatory publishing in academia. Predatory publishing is a term popularized by Jeffrey Beall to refer to journals that charge huge fees to submit papers without proper peer review. Predatory publishing damage the scientific process by cheapening intellectual work and misleading scholars, especially early career researchers.

Beall created the list in 2008 [64], and from 2010 to 2014 alone the size of the list increased ten-fold, growing to include thousands of journals and publishers [65]. Inclusion on the list was based on a metric derived from a 52-point criterion that Beall created himself [66].

The list was controversial, mostly due to Beall's biases and previous positions against the open-access movement he described as "anti-corporatist, oppressive and negative" [67], or strong statements in the lines of "predatory publishing damages science more than anything else" [68]. Despite the controversy, evidence points to the fact that Beall's list highlighted recognized problems in academia, and set to worsen [65] [69]. Other studies point to the additional fact that the issue is strongly regional, and expected to worsen even further, as scientific research turns into a global endeavor [70] [71] [65].

On January 15th of 2017, Beall took his site and the list down, due to "threats and politics" [72].

Private initiatives swiftly took on to seize the opportunity and fill the void [73] through centrally managed "black" and "white" lists. As it is usually the case with centrally managed initiatives, it ignores "local knowledge"[19] [74] and fails to address the causes of the negative phenomena[20].

---

[18] The published author of the paper "D. H. J. Polymath" is a reference to the proposed problem, a new proof of the Density Hales-Jewett theorem, and to the crowd that took part in the resolution during the Polymath project

[19] The "local knowledge problem" in economics is often used to explain why the central control of distributed resources (including centrally planned economies) does not work [74]

[20] Even if unintended, there is symbiotic relationship in place - the very existence of a centrally managed, subscription based list is justified by the

At that point, a community-based initiative, called "Stop Predatory Journals", ran by an anonymous community, took on the maintenance of the original list and extended it. The initiative mostly gathers contributions made through a simple configuration management platform and keeps a publicly available list of predatory journals, predatory publishers, hijacked journals[21], and misleading or fake metrics.

Despite a positive impact, this community-based initiative is still open to criticism, but more objectively, a crowd-based initiative has to look primarily at market incentives in large scale. In this sense, predatory publishing can be seen as a market-oriented, rational response to two factors:

- A poor system of incentives currently in place in academia [75];
- Bad funding models. There should be more than 'author pays' or 'reader pays' models. The actual cost of publication is a fraction of what used to be when these systems were designed. Additionally, other financial costs like peer review are very relative in an environment that relies on a system of incentives for virtual collaboration [76].

In closing, the current status of crowd-based surveillance of predatory journals is positive but fail to address the root causes of the phenomena. The overall solution lacks adequate computational representation for academic content and aligned system incentives for collaboration, considering the requirements for large-scale collaboration introduced in Section II.

In general, empirical evidence shows that collaborative crowds are more appropriate for the resolution of complex problems than conventional methods. Current research, however, is not able to pinpoint the exact reasons, or characteristics, of problems that would be more suitable for resolution by collaborative crowds [77] [78]. With a few exceptions, current literature lacks a quantitative analysis of the suitability of crowds for the resolution of complex problems [79].

## V. FROM POLYMATHS TO CROWDS: A HISTORICAL PERSPECTIVE

It should come without surprise to most people that the ability to build objective knowledge through a scientific method is what drove humans out of caves and shot our race towards the stars. As we have previously explained in Section I, the scientific method can be seen as a cumulative process in a sense that, over time, we build new knowledge based on previous knowledge considered to be true. Truth, or at least what we perceive to be true, is not constant [21]. Given our history of understanding of the world around us, previous knowledge will almost certainly be ruled as false at some time in the future. On this erratic pathway a "tapestry" of "knowns" slowly evolves to take the infinite space of "unknowns" based on ever-changing knowledge foundations [80].

---

existence of the damaging practice - that serves as a reverse incentive to ending the practice of predatory publishing altogether.

[21] A hijacked journal is a journal that had either their websites or branding co-opted by a predatory journal or publisher.

This dynamic and seemly chaotic method of acquiring an understanding of the world around us has been happening for as long as our ancestors started to grasp with inquiries and guesses about cause and effect of observable phenomena. Even if it was unintended, and we were not entirely aware of its exact mechanisms, this organic adaptation has been happening, constantly. It is so ingenious and so ancient that it has been organically adjusting itself to an ever-changing knowledge base, resources, and culture available at different points in history and time.

This adjustment occurs in evolutionary stages in response to available technology, the individual performing the scientific investigation, drivers, collaboration, creativity, and control in three distinct phases, as described in Figure 3.

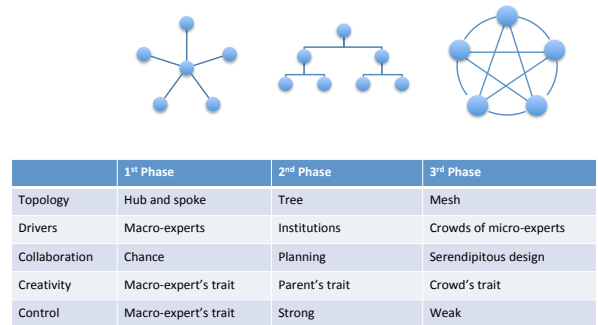

|  | 1st Phase | 2nd Phase | 3rd Phase |
|---|---|---|---|
| Topology | Hub and spoke | Tree | Mesh |
| Drivers | Macro-experts | Institutions | Crowds of micro-experts |
| Collaboration | Chance | Planning | Serendipitous design |
| Creativity | Macro-expert's trait | Parent's trait | Crowd's trait |
| Control | Macro-expert's trait | Strong | Weak |

**Figure 3. Phases of Collaborative Scientific Investigation**

Characteristics of the evolution of collaborative scientific investigation from macro-experts, to institutions, and to crowds, depending on five factors: the individual performing the investigation, drivers, collaboration, creativity, and control

This chart summarizes each of the phases according to the individual performing the research, topology, drivers, collaboration, creativity, and control. The individual performing the investigation evolved from macro-experts, or "natural philosophers", to groups arranged hierarchically organized in institutions, to an upcoming phase in which micro-experts are arranged in mesh-like crowds, subject to the requirements listed in Section II. Each of the topologies represents the communication paths between participants in each phase. The drivers for discovery, or the entity in charge of conducting the investigation and inquiry, in each phase, are macro-experts, institutions, and crowds of micro-experts. Collaboration occurs in each of the phases by chance, central planning, and by serendipitous design, as explained previously in Section III. Creativity in each phase is associated with the individual performing the investigation. Finally, control is either trait depending on the macro-expert, strong, or weak, depending on the topology in place, respectively hub-spoke, tree, or mesh.

The first phase of scientific investigation relied uniquely on "natural philosophers", bright individuals who were able to drive discovery based on their personal traits and occasional interaction with other "natural philosophers".

During this first phase, the domain of investigation was related to natural observations and research conducted by individuals almost in isolation. Given the relative simplicity of subjects under study, a few very bright individuals could still build on previous knowledge with little or no interaction with other individuals. The collaboration was done on an ad hoc basis, and opportunities for interaction were left to chance and rare social exchanges. The proximity with domains of study would allow for self-funding, and the management of resources is de-centralized and done in almost complete isolation. Ultimately, the expertise held by any single individual would determine the effectiveness of one's research – this is the golden age of polymaths or *macro-experts*.

Even with all of these intrinsic limitations, objective learning occurred, and the accumulation of knowledge led to increased complexity and a higher demand for resources to record, store and share knowledge. The ever-increasing body of knowledge demanded a more significant interaction with other researchers that would not necessarily share the immediate surroundings where research was taking place. Large institutions came along to fulfill the demand and manage the vast amount of resources needed for more complex methods, and more information.

That triggered to the second phase, in which institutions took over the task of organization of participants and the management of resources required for investigation and collaboration. As more and more resources were needed as multidisciplinary subjects increased in complexity, this second wave, the phase of institutionalized science, came to life.

In this second phase, participants were organized in hierarchical institutions across diverse kinds of institutions, interested in or dependent on scientific advancement: academia, governments, or private corporations. Most scientific procedures evolved to match the hierarchical organization of these institutions, and so the production of objective knowledge followed.

## VI. Disruption and Breakthrough

As we move along through the second historical phase, institutionalized science started to shape research methods to fit into a more cumbersome, hierarchical communication, relying on larger, less efficient group sizes. Large institutions also brought along the unintended consequence of heavy top-down hierarchical communication and stronger controls. The immediate consequence overall was that the complexity of research domains started to increase exponentially.

This increased complexity has been producing two major changing forces that are shaping the resurgence of a next historical phase:

- **Multidisciplinary collaboration**: multidisciplinary collaboration became mandatory. We cannot perform an objective investigation, on any field, without an understanding of orthogonal fields of knowledge. Unlike the first phase, no single participant, regardless of how bright, detains enough expertise to provide a full, overreaching solution to a modern-day problem. The natural limitation of individual participants of the scientific process in dealing with an ever-growing knowledge body marks the beginning of the demise of the age of macro-experts.

- **Complexity requires control**: technology is an amplifier of features present in any environment, regardless of how we perceive the results of these features as positive or negative. This amplifying side effect of technology is observable everywhere: in politics, personal and business relationships, financial markets, and especially in our topic of interest: scientific procedures applied to economics. Technology is getting to a level of complexity and sophistication that its scientific use without control plays a role of a double-edged sword: it can cause more harm to the development of objective knowledge than good. One should expect the same scientific method that brought significant technological advancements would naturally improve the tools available for investigation, specifically computational tools. It did so to a certain extent. It is true we have advanced technology and methods available in the scientific investigation, but it is also true that we have abundant evidence of misuse of computational resources and methods in the scientific investigation leading to wrong or corrupt data and as a consequence defective research.

These two forces are bringing several disruptive manifestations as signs of an upcoming wave of transformation. These manifestations, listed over the next paragraphs, are evidence that this second historical phase of the scientific investigation is presenting signs of inadequacy with current status of technology, historical, and cultural developments:

- **Misaligned Academic Incentives:** despite an organized and commendable effort by scientific institutions to contain this harmful practice, evidence shows that current academic incentives are fostering a culture of fraud. Based on pools and questionnaires, research finds an astonishing number of scientists engaging in a range of behaviors extending far beyond falsification, fabrication, and plagiarism [81]. The issue is so prevalent that quantitative models can reliably predict and estimate the number of articles that should be retracted over time [82].

- **Hierarchies Stifles Creativity:** scientific research is mostly driven by creativity. As explained earlier in this section, in the current phase of scientific investigation, work is often performed in hierarchical structures. While hierarchies work reasonably well for control and decisions, there is evidence that hierarchies are detrimental to creativity [83] [84]. There is also evidence that while hierarchical institutions usually verbally request for innovation, their top-down structures unintentionally reject them [85] [86] [87]. The ideal structure to foster creativity is closer to a peer-to-peer association, the one presented by human crowds, than a top-down hierarchical structure [88].

- **Human Limitation on Information Processing:** there are hard limitations on how much information humans can process. The limitation on how much information we can consume and understand also limits the throughput of quality research scientists can produce, review and reproduce [89]. There is evidence that scientists have already reached a plateau on how much information they can efficiently absorb, handle, and produce [90].

- **Unavailability of Quality Academic Content:** major journals have recently placed restrictions on mining and use of scientific data in large scale [38]. Similar limitations apply to the refusal of providing details on landmark research findings for "reasons of confidentiality" [91] [92] [93]. These restrictions undermine both the automation of hypothesis generation reliant on vast amounts of quality data and crowd collaboration of micro-experts dependent on access to peer-reviewed, quality academic content.

- **Lack of Means to Record and Share Reliable Data:** lack of appropriate means to record and share reliable data has been indicated as one of the limiting factors in modern investigation procedures. An additional limiting factor is the lack of a central authority to validate observations and a central repository of evidence-based knowledge [94]. Other evidence in the field of economics describe examples of global economic policies defined based on flawed data stored in plain excel spreadsheets [95] [96] [97].

- **Science Hacking:** evidence collected in a correlated field show a considerable rate of complex biotech experiments published in prominent journals, heavily reliant on advanced computational resources, just cannot be appropriately reproduced [98]. Additionally, evidence shows that reproducibility is negatively correlated with the relative computational complexity of the experiment. In the field of economics, in particular, we have similar evidence [99] measuring that only 61% of the articles in a major journal of economics can be successfully reproduced. Similar results have been found in psychology, in which only 38% of the studies can be successfully reproduced [100], or biotechnology where only 6 out of 53 "landmark cancer studies", i.e., 11%, could be properly reproduced [93]. Quantitative metrics also show examples of "statistics used wrong", proliferating the belief that p-values alone can determine findings to the true when in reality they are false [101] [102].

Modern science and the associated scientific method have taken a critical role in human societies. We have learned to blindly trust lives and outcomes of global reaching economic policies to findings that should be shielded from scrutiny by merely labeling them 'scientific'. If these manifestations listed above sound alarmist, the feeling is rooted in plausible reasons.

Now, here comes time for the third phase of collaborative scientific investigation based on multidisciplinary, diverse collaboration in large-scale through crowds. The use of a crowd-based investigation to serve as an attenuator of the changing forces disrupting institutionalized science.

## VII. Conclusion

Increased complexity is imposing on researchers a mandatory multidisciplinary thinking and the use of advanced computational technology that is too advanced for most people, even scientists, to properly use and understand. The consequence, a disproportional amount of scientific results cannot be reproduced.

To counteract this deviation, we have to act on two fronts: first, how to make macro-experts collaborate properly, and second by controlling evidence of computational artifacts of any kind – e.g., data, models, plots – so they can be adequately understood, investigated, traced and replicated.

This paper relates to the use of technology for improvement of methods of investigation on these two fronts: proper collaboration and computational controls. These topics cover a broad variety of related academic work, defined by two ends of a spectrum, as shown in Figure 4.

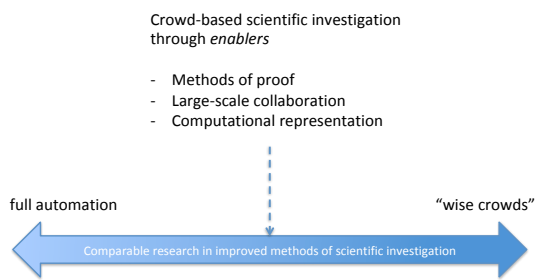

**Figure 4. Comparable Research in Methods of Scientific Investigation**

The spectrum defining the topic of improved methods of scientific investigation covers a broad range. On the right end of the spectrum are methods that rely on the full automation of methods of investigation. On the opposite end, methods relying on "wise crowds". This research, proposing the use of specific enablers of a crowd-based investigation, sits somewhere in-between the two opposing ends.

One end of the spectrum is defined by research that intends the full automation of the process of investigation. This approach relates to the assignment of computers to execute tasks that are usually performed by scientists. Some of those tasks are associated with the registration of observations [29], automated hypothesis generation [32] and contextual gaps [33], and automated testing based on robotics [41]. Fully automated research is not feasible given current technology, as previously discussed in the definition of the proof pipeline, in Section II. Research on fully automated methods of investigation usually brings the same consequences and criticisms associated with data-driven research [103], in which correlations of data are detected first, and only then, hypotheses are produced.

On the other end of the spectrum are solutions that rely on the existence of "wise-crowds", based on empirical evidence collected over the years [51] [52] [53]. The wise-crowds approach assumes the existence of some invisible, unquantifiable mechanism that makes crowds wise, and relies on the assumption of complete independence and decentralization. Paradoxically, the assumption of independence would diminish the value of structured collaboration in crowd investigation. Evidence on the existence of some mechanism enabling wise-crowds to occur is often empirical [104] and the subject of some criticism [105].

This research sits somewhere in the middle of this spectrum. We advocate the use of a crowd-based investigation through methods of proof, large-scale collaboration, and computational controls. This research emphasizes the use of computers for mechanical and repetitive tasks, like the orchestration of scientific interaction in crowds and the record of scientific evidence as contributions, as previously described in Section II. Additionally, this research also advocates for the importance of intangible human factors related to experience and creative thinking in science, and a hypothesis-driven process of discovery.

This next generation investigative approach is an organic evolution of how scientific participants have been interacting given resources available at the time. From polymaths, to a centralized institutionalized science, to an upcoming form of crowd-based, distributed science. This paper advocates leveraging large-scale collaboration as a method of self-organization for investigation in complex fields of knowledge.

We expect this change to be disruptive and to be far from contained in academia only. The effects of the crowd automation in the acquisition of objective knowledge will be reflected in human interactions on all levels, and on a global scale.